\documentclass[preprint,amsmath,amssymb]{revtex4}
\usepackage{graphicx}% Include figure files
\usepackage{bm}
\begin{document}

\title{Communication activity in a social network: relation between
  long-term correlations and inter-event clustering}

\author{Diego Rybski$^{1,2}$, Sergey V. Buldyrev$^{3}$, Shlomo
  Havlin$^{4}$, Fredrik Liljeros$^{5,6}$, and Hern\'an A. Makse$^{1}$}

\affiliation{
  $^1$ Levich Institute and Physics Department, City
  College of New York, New York, NY 10031, USA\\
  $^2$ Potsdam Institute for Climate Impact Research (PIK), 
  P.O. Box 60 12 03, 14412 Potsdam, Germany\\
  $^3$ Department of Physics, Yeshiva University, 
  New York, NY 10033, USA\\
  $^4$ Minerva Center and Physics Department, 
  Bar-Ilan University, Ramat Gan 52900, Israel\\
  $^5$ Department of Sociology, Stockholm University, 
  S-10691 Stockholm, Sweden\\
  $^6$ Institute for Futures Studies, Box 591, 
  SE-101 31 Stockholm, Sweden\\}

\date{\today \enspace -- \jobname}

\begin{abstract}

The timing patterns of human communication in social networks is not
random. On the contrary, communication is dominated by emergent
statistical laws such as non-trivial correlations and clustering.
Recently, we found long-term correlations in the user's activity in
social communities. Here, we extend this work to study collective
behavior of the whole community. The goal is to understand the origin
of clustering and long-term persistence. At the individual level, we
find that the correlations in activity are a byproduct of the
clustering expressed in the power-law distribution of inter-event
times of single users. On the contrary, the activity of the whole
community presents long-term correlations that are a true emergent
property of the system, i.e. they are not related to the distribution
of inter-event times. This result suggests the existence of collective
behavior, possible arising from nontrivial communication patterns
through the embedding social network.

\end{abstract}
%Communication can be considered as the exchange or transmission of
%information.  Embedded in a social network, the communicating
%partners communicate in a complex fashion, where the act of
%communication is driven by complex internal and external influences.
%However, the origins of this long-term memory are unclear: From a
%statistical physics point of view, long-term correlations can be due
%to (i) power-law distributed inter-event times (Levy correlations) or
%(ii) long-term memory between the activity at different times.  Here
%we study the possible origin of the persistence where we find
%evidence for Levy correlations at the single user level.  However,

%\noindent{\it Keywords\/}:
%human dynamics, communication, long-term persistence, Levy correlations

%\pacs{64.60.aq, 89.65.-s, 89.70.-a, 89.70.Hj, 89.75.Da, 95.75.Wx}

%\submitto{Journal of Statistical Mechanics}

\maketitle

Various constituents of social systems have been found to follow
remarkable statistical regularities.  Only the recent availability of
relevant data made it possible to unravel such features.  Tracking
bank notes or cell phones it has been shown that humans follow simple
and reproducible mobility patterns
\cite{BrockmannHG2006,GonzalezHB2008}.  
%Sexual networks were found to
%be scale-free, i.e. comprising a power-law distribution of the number
%of sexual partners \cite{LiljerosEASA2001}.  
The communication via
e-mails occurs in bursts, exhibiting a broad distribution of times
between successive messages of individuals (inter-event times)
\cite{BarabasiAL2005,MalmgrenSMA2008}.  Recently, we have found that
the act of sending messages of individual users in two online
communities present long-term correlations
\cite{RybskiBHLM2009} characterized by power-law correlation functions
obtained via standard Detrended Fluctuation Analysis.

  In the present work we examine the relation between the two
  empirical findings of broad inter-event time distributions
  \cite{BarabasiAL2005,MalmgrenSMA2008} and the long-term persistence
  identified in the communication activity \cite{RybskiBHLM2009}.
  Therefore, we investigate the communication activity of actants in a
  social online community with special consideration of the timing and
  study long-term correlations in the communication as well as
  clustering of successive messages.

Long-term correlations have been found in the dynamics of many
physical, technological, and natural systems.  They are characterized
by a divergent correlation time, i.e. a power-law decaying
auto-correlation function (for a review see \cite{KantelhardtJW2010}).
Such correlations lead to a pronounced mountain-valley-structure on
all time scales -- comprising indeterministic epochs of small and
large values \cite{makse-pre}.  This type of persistence represents a
surprising regularity since it is present in many different data such
as DNA-sequences, human heartbeat, climatological temperature, etc.
\cite{PengBGHSSS1992,PengMHHSG1993,KoscielnyBundeBHRGS98}.  Long-term
persistence in human related data has been reported for highway
traffic \cite{TadakiKNNSSY2006,XiaoYanZHM2007}, Wikipedia access
\cite{KaempfTKM2011}, Ethernet traffic \cite{LelandTWW1994}, finance
and economy \cite{LiuGCMPS1999,MantegnaS1999,LuxA2002}, written
language \cite{SchenkelZZ1993,KosmidisKA2006}, as well as
physiological records
\cite{PengMHHSG1993,IvanovBAHFBSG1999,BundeHKPPV00}, Human brain
activity \cite{LinkenkaerHansenNPI2001,AllegriniMBFGGWP2009,brain} and
human motor activity \cite{IvanovHHSS2007} also comprise long-term
correlations as well as city growth
\cite{makse-city,makse-perco,rozenfeld-pnas,rozenfeld-aer}, biological
networks \cite{galvao} and the spreading of disease \cite{obesity}.

% Recently, long-term correlations have been found in online posts by
%Brazilian sex sellers and buyers \cite{RochaH2010}.

The distributions of inter-event times (times between successive
messages)
%(times between successive messages of individual persons)
have been found to be rather broad, described by power-laws
\cite{BarabasiAL2005}.  If many short intervals are
separated by few long ones, the activity as messages per unit time
comprises persistence, i.e. epochs of large and small activity.  Since
such distributions have been described with power-laws, we wish to
investigate the relation between the long-term correlations in
activity \cite{RybskiBHLM2009} and the broad (power-law) distribution
of inter-event times \cite{BarabasiAL2005}.  We will
test two possible scenarios: (i) In the first scenario, the long-term
correlations found in the communication activity \cite{RybskiBHLM2009}
result from Levy type distributions, i.e. correlations are only due to
the power-law inter-event time distribution (with exponents in the
specific range) \cite{ShlesingerFK1987}.  In the second scenario, (ii)
the activity comprises 'real' correlations, i.e. the inter-event time
distributions do not follow a power-law, but the communication
activity is temporally not independent, namely long-term correlated.

We study the activity of sending messages based on detailed temporal
data from a social online community and obtain the long-term
correlation exponent $H$ via DFA.  The exponent $H$ depends on the
overall activity of the members; the more active the members the
larger the fluctuation exponents. This exponents reaches a value
$H\approx 0.90$ for the most active users from an uncorrelated value
$H\approx 0.5$ for the less active ones. Then, we compare the value of
$H$ with the corresponding exponents of randomized data and a
theoretical prediction relating correlations with clustering in the
inter-event times.  From the consistency of the comparison of this
three measures, we conclude that the long-term correlations found in
the activity of sending messages for single users is a direct
consequence of the power-law distributed inter-event time of the
individuals. Thus, the burstiness in the user activity explains the
long-range correlations.

More interesting results are found when we consider the activity of
the whole community as a sum of the activity of its members. Again we
find non-trivial long-range correlations with exponents $H$ in the
same range as the individual users. However, the origin of this
correlations is not related to the inter-event activity. This is
probed by shuffling the activity data but preserving the distribution
of inter-event times. In this case, this shuffling destroys the
long-range correlations, implying that the correlations are not a
byproduct of the broad distribution of inter-event times. We conclude
that the whole system acts as a true long-term correlated system where
correlations are not directly related to the Levy distributions of
events.

%This paper is organized as follows.  In the data section we briefly
%describe the online community where the communication activity was
%logged and which information the data includes.  In
%Sec.~\ref{sec:analysis} we present the performed analysis and describe
%our results.  We discuss and draw conclusions in
%Sec.~\ref{sec:conclusions}.

%\subsection{Data}
%\label{sec:data}

We analyze the data of an online community (www.pussokram.com, POK,
\cite{HolmeP2003,HolmeLEK2003,HolmeEL2004})
covering the complete lifetime of the community over $492$\,days from
February 2001 until June 2002. We record the activity among almost
$30,000$ members with more than 500,000 messages sent.  This
internet-site has been used for general social interactions and
dating.  The data consist of the time when the messages are sent and
anonymous identification numbers of the senders and receivers.  The
data has been analyzed by us in \cite{RybskiBHLM2009,RybskiBHLM5b}.
%There is no message content.  
In contrast to similar network data sets
consisting only of snapshots, i.e. temporally aggregated social
networks expressing who sent messages to whom, the advantage of this
data set is that it provides the exact time when the messages were
sent.  For a discussion see \cite{GallosRLHM2011}.

Before shutdown, the members could log in and meet virtually.  In such
communities, there are different ways of interacting.  Usually, it is
possible to choose favorites, i.e. certain members, that a person
somehow feels committed to.  Such platforms also offer the possibility
to discuss in groups with other members about specific topics.  We focus
on messages sent among the members -- they are similar to e-mails but
have the advantage that they are sent within a closed community where
there are no messages coming from or going outside.
Figure~\ref{fig:pokillu} illustrates patterns of sending
messages for typical single users [a-d] and for the whole community
[e].  The data is publically
available\footnote[0]{http://lev.ccny.cuny.edu/$\sim$hmakse/soft\_data.html}.
We would like to note that we do not consider here the QX dataset
which we analyzed in \cite{RybskiBHLM2009,RybskiBHLM5b}, since it
covers only $2$~months and the scaling of the distribution of
inter-event times is not reliable and we could not measure the shape of
this distribution consistently.

% 80,683 members
% 12,590,896 messages
% 2,172,185 directed
% 1,370,730 undirected
% 528,098 messages
% 28,876 members
% 174,151 directed
% 115,324 undirected

\section*{Results}
\label{sec:analysis}

\subsection*{Study of correlations in individual activity}

Applying DFA \cite{PengBHSSG94,BundeHKPPV00,KantelhardtKRHB01} we have found in
\cite{RybskiBHLM2009,RybskiBHLM5b} that the individual activity
records, $x(t)$, i.e. messages per unit time (records of messages per
day or per week), exhibit long-term correlations. The fluctuation
function provided by DFA scales as
\begin{equation}
F(\Delta t)\sim (\Delta t)^H,
\label{dfa}
\end{equation}
where the exponent~$H$ is also known as the Hurst exponent.  In the
case of long-term correlations -- which are characterized by a
power-law decaying auto-correlation function:
\begin{eqnarray}
C(\Delta t) & = & \frac{1}{\sigma_x^2}\left\langle
\left[x(t)-\langle x(t)\rangle\right]
\left[x(t+\Delta t)-\langle x(t)\rangle\right] 
\right\rangle \nonumber \\
& \sim & (\Delta t)^{-\gamma} \nonumber
\, ,
\end{eqnarray}
where~$\langle \cdot \rangle$ denotes the average, $\sigma_x$ is
standard deviation of $x(t)$, and $\gamma$ is the correlation exponent
($0\le\gamma\le 1$) -- one finds $1/2\le H\le 1$, whereas larger
exponents correspond to more pronounced long-term correlations.  For
uncorrelated or short-term correlated records ($\gamma \ge 1$, or in
general $\gamma \ge d$, $d$ is the substrate dimension) the asymptotic
fluctuation exponent is $H=1/2$.  In the range $0 \le \gamma \le 1$
both exponents are related via
\begin{equation}
\label{eq:h1g2}
\gamma=2-2H
\, .
\end{equation}
For an overview, we refer to \cite{KantelhardtKRHB01,KantelhardtJW2010}.
DFA$n$ removes polynomial trends of the order $n-1$ from the original 
record~$x(t)$, i.e. DFA2 copes with linear trends.

It is important to note that the DFA fluctuation function
Eq.~(\ref{dfa}) is not applied to the activity $x(t)$, but to the
integrated signal $y(t) = \sum^t x(t')$. Thus $x(t)$ would be the
analogous to the steps in a random walk and $y(t)$ the
displacement. DFA incorporates an additional detrending of the
data. The integration leads to the appearance of long range correlation
when the interval between each step is power-law distributed. We will
come back to this result when explaining the long-term correlations in
terms of the burstiness.

We have measured the fluctuation exponents by applying least squares
fits to $\log F(\Delta t)$ vs. $\log \Delta t$ on the scales
$10<\Delta t<70$\,weeks conditional to the member's activity level,
e.g. their total number of messages, $M$ \cite{RybskiBHLM2009}.
Figure~\ref{fig:dfaaveallbinfitiet} depicts the DFA results. We find
that the less active members, sending only one or two messages in the
period of data acquisition, exhibit uncorrelated behavior.  The more
messages the members send, the more correlated is their activity.  The
fluctuation exponent~$H$ increases with~$M$ and reaches values up to
$H=0.91 \pm 0.04$ (value obtained for sending messages, we disregard
the last points, $M>400$, which have too large errors bars).  The
uncorrelated behavior-- $H\approx 0.5$-- for small
activity can be understood since when $M\approx 1-10$ there is
not enough time in the data acquisition window to capture long-term
correlations.  In \cite{RybskiBHLM5b} we propose a model which
reproduces the dependence of the fluctuation exponents on the activity
level of the members.  For receiving messages we find almost identical
results \cite{RybskiBHLM5b}.  We use weekly resolution in order to
cope with possible weekly oscillations
\cite{GolderWH2006,LeskovecH2008,MalmgrenSMA2008,MalmgrenSCA2009}.

Similar long-term correlations have been found in
\cite{EislerK2006,EislerBK2008} in traded values of stocks and e-mail
communication. The fluctuation exponent increases with the mean
trading activity of the corresponding stock or with the average number
of e-mails similarly as in our results.

\subsection*{Study of clustering in individual activity}

The timing of human communication activity has been found to comprise
bursts where many events occur in relatively short periods which are
separated by long periods with few or no events at all.  Such patterns
can be characterized with the inter-event times, i.e. the times, ${\rm
  d}t$, between successive messages.
%i.e. the times, ${\rm d}t$, between successive messages of individuals. 
For e-mail communication it has been argued that their probability density 
follows a power-law, 
\begin{equation}
\label{eq:pdtdelta}
P({\rm d}t)\sim ({\rm d}t)^{-\mu}
\, , 
\end{equation}
with exponent $\mu\approx 1$
\cite{BarabasiAL2005,JohansenA2004,JohansenA2006}.  As an origin for
such heavy tails in human dynamics a queuing model has been suggested
\cite{BarabasiAL2005} according to which each individual performs
tasks from a priority list.  It has been confirmed that such a process
can reproduce bursts of activity or clustering, see
e.g. \cite{VazquezA2005,VazquezODGKB2006}.  In contrast, analyzing the
same e-mail data, a log-normal distribution has been found to be more
appropriate to describe the inter-event time distribution
\cite{StoufferMA2005,BarabasiGV2005}.  We would like to remark that
fitting fat tailed distributions is disputed
\cite{NewmanM2005,GabaixI2006,ClausetRSN2007,MalevergnePS2009}.  There
is neither a consensus on a typical functional form nor on a proper
fitting technique.  Recently, a cascading Poisson process based on
daily and weekly cycles has been proposed as origin of
slower-than-exponential decays of $P({\rm d}t)$
\cite{MalmgrenSMA2008,MalmgrenSCA2009}.  We studied the cascading
Poisson process in \cite{RybskiBHLM5b}.

In \cite{GohB2008}, memory in the sequences of ${\rm d}t$ has been studied 
for different data sets, characterizing the inter-event times in terms of a 
burstiness parameter, which is based on the distribution, 
and in terms of a memory coefficient, 
which is the auto-correlation function at lag~$1$. 
In addition, the authors locate the corresponding data sets in a 
phase diagram defined by these two quantities. 
Nevertheless, we would like to note that the quantification of 
long-term correlations in the ${\rm d}t$ can be hindered by 
noise \cite{EichnerKBH2007,LennartzB2009}.

Next, we study the POK data, i.e. the inter-event times~${\rm d}t$ 
between successive messages of individual members, 
and relate their statistics to the long-term correlations.
The finding of long-term correlations opens the question of the origin of 
such a persistence pattern in the social communication. 
From a statistical physics point of view, 
we consider two possible scenarios: 
\begin{enumerate}
\item
In the first scenario, the intervals between the messages follow a
power-law \cite{BarabasiAL2005,GersteinM1964}.  Accordingly, the
activity pattern comprises many short intervals and few long ones,
implying persistent epochs of small and large activity.  This
fractal-like clustering in the activity can -- depending on the
exponent -- lead to long-term correlations with $H>1/2$ (see the
analogous problem of the origin of long-term correlations in DNA
sequences as discussed in \cite{BuldyrevGHPSS1993}).  This scenario
implies a direct link between the correlations in the activity and the
distribution of inter-event times which can be obtained analytically
\cite{BuldeyrevSV2010}. We call this scenario ``Levy correlations''
since the actual activity may not be correlated per-se, but
correlations arise as a byproduct of integrating a signal with a
power-law distribution of inter-events in the DFA formalism.

\item
In the second scenario, the intervals between the messages may or may
not follow a power-law distribution, but the values of the inter-event
times are not independent of each other and comprise 'real' long-term
persistence.  For example, the distribution of inter-event times could
be stretched exponential (see recent work on the study of extreme
events of climatological records exhibiting long-term correlations
\cite{BundeEKH2005,EichnerKBH2007}) and then the only way to explain
long-term correlations in the activity are correlations in the
inter-event times. We call this scenario ``true correlations'' since
the correlations are not related to the distribution of inter-events
but they reflect 'real' correlations in the dynamics of the
communication activity.
\end{enumerate}

A possible way to discern between these two scenarios is to shuffle
the temporal activity, keeping the inter-event distribution
intact.  While in the case of Levy type correlations shuffling the
inter-event times should not influence the long-term correlation
properties of $x(t)$, in the case of 'real' long-term correlations
shuffling the inter-event times should destroy the (asymptotic)
long-term correlations since the memory is due to the arrangement of
the inter-event times.  In what follows, we investigate the activity
of individual members and the activity of the whole POK community.

\subsection*{Study of inter-event distribution of individual members}

Figure~\ref{fig:dfaaveallbinfitiet} exhibits the fluctuation exponents
for individual members when we shuffle the data but preserve the
distribution of inter-event times.  The corresponding exponents also
reach high values, almost as high as for the original data, and do not
drop for very active members.  This agreement is a first
indication of Levy correlations in single user activity.

Further evidence is found by studying the distribution of inter-event
times in the activity of each individual.  Figure~\ref{fig:ietall}
shows the probability density, $P({\rm d}t)$, of times between
messages of the same users sent in the online community.  A power-law regime of
approximately two decades can be seen with an exponent $\mu\approx
1.5$, which differs from the exponent reported for e-mail
communication \cite{BarabasiAL2005,JohansenA2006}, i.e. $\mu\approx
1$.  A reason for these different findings might be that in the case
of \cite{BarabasiAL2005} only one user is considered and that $\mu$
depends on the activity level of the users, as we show below.  In
addition, here we study all messages from a closed community.  The
exponent we find is closer to the one reported for reply times
(waiting times), i.e. the time individuals spend between receiving and
sending to the same communication partner.  For reply times of e-mails
and land mail $\mu_{\rm w}\approx 1.5$ has been reported
\cite{BarabasiAL2005,OliveiraB2005}.

Since we found a dependence of the fluctuation exponent~$H$ on the
activity level~$M$, i.e. the total number of messages each member
sends, we suspect that also~$\mu$ might depend on~$M$.  Thus, in
Fig.~\ref{fig:oc2iet} we plot for sending messages in POK (daily
resolution) the $P({\rm d}t)$ for groups of different activities,
i.e. different total number of messages~$M$.  We find that for the
most active members $P({\rm d}t)$ decays rather steeply, while for the
least active members $P({\rm d}t)$ decays much slowly.  Due to the
finite size of the data it is not quite clear which functional form
the curves follow.  If one assumes a power-law decay then the
exponents are roughly in the range $1\le\mu\le 3$.

As discussed above, the power-law distribution of inter-event times,
Eq.~(\ref{eq:pdtdelta}), can lead to long-term correlations in
activity, without requiring temporal dependencies between the
intervals themselves. It can be shown that the long-term persistence
properties of this point process are characterized by the fluctuation
exponent which theoretically depends on $\mu$ according to
\cite{ShlesingerFK1987,ThurnerLFHFT1997,AllegriniMBFGGWP2009,BuldeyrevSV2010}:
\begin{equation}
\label{eq:hdelta}
H_\mu=
\left\{\begin{array}{cl} 
\mu/2 & \mbox{ for }1<\mu<2\\ 
2-\mu/2 & \mbox{ for }2<\mu<3\\
1/2 & \mbox{ else}
\end{array}\right.
\, ,
\end{equation}
see Fig.~\ref{fig:prediction}.  Apart from detrending, DFA provides an
integration of the original record.  So if there are long periods of
no activity due to power-law inter-event times, then, this is
reflected in long persistence in the signal calculated by DFA.  Thus,
the existence of long-term correlations is due to the long periods
distributed via Levy distributions as expressed by the direct relation
between correlations and Levy inter-event activity,
Eq.~(\ref{eq:hdelta}).

Applying least squares fits (in the straight range) to the $P({\rm
  d}t)$ for sending in POK (Fig.~\ref{fig:oc2iet}) we obtain values
for $\mu$ as a function of the activity level~$M$ and determine the
corresponding fluctuation exponents, $H_\mu$, as expected from
Eq.~(\ref{eq:hdelta}).  We would like to note that the curves in
Fig.~\ref{fig:oc2iet} are not always straight lines leading to large
uncertainty regarding the estimated values of $\mu$.

Figure~\ref{fig:dfaaveallbinfitiet} depicts the fluctuation
exponents $H_\mu$ from Eq.~(\ref{eq:hdelta}) in comparison with the
values obtained from DFA.  We find $H\approx H_\mu$ for a big part of
the $M$ range. The exponents $H_\mu$ are also close to $H$ of the
shuffled records where the inter-event times are preserved. The fact
that when we shuffle the signal, respecting the corresponding
distribution of inter-event, gives rise to the same correlation
function, indicates that the origin of the long-term correlation
obtained in DFA are due to the Levy correlations. This is further
corroborated by the agreement between $H$ from DFA and the prediction
$H_\mu$.  From Fig.~\ref{fig:dfaaveallbinfitiet} we see that the
three curves are in a reasonable agreement.  This supports that the
correlations in single user activity can be due to the power-law
distribution of the inter-event times, which is in favor of Levy type
correlations.

\subsection*{Study of whole community activity}

Next, we investigate the activity of the community as a whole. While
we have studied the activity of single users, it is of interest to
investigate the activity of the whole community by considering the
number of messages sent by all members in a specified period of time.
Figure~\ref{fig:pokillu}(e) shows such activity temporally aggregated
to one day. The interest arises since we would like to test the
existence of correlations arising from collective behavior in the
communication patterns at the level of the whole community.

For this study, we disregard who sends the messages to whom and only
consider the instants when any message was sent.  In order to have a
sufficiently long record to apply DFA, we aggregate the data to
messages per hour (instead of daily or weekly resolution).  As can be
seen in Fig.~\ref{fig:pokillu}(e), the record contains oscillations
\cite{MalmgrenSMA2008}.  Since such periodicities lead to erratic
fluctuation functions \cite{KantelhardtKRHB01}, we subtract the hourly
averages over all days: $x_{\rm tot}(t) \rightarrow x_{\rm tot}(t)-
\langle x_{\rm tot}\rangle_{t\,{\rm mod}\,24}$.

The DFA fluctuation functions are shown in
Fig.~\ref{fig:pokrecall3600seadfa}.  The hump on scales around $20$
hours in the results of DFA1 and DFA2 are residual oscillations,
i.e. they were not completely removed.  On larger scales this effect
vanishes and we find a fluctuation exponent $H_{\rm tot}\approx
0.9$. The straight line in the case of DFA0 is due to the fact that
the maximum exponent is 1 \cite{KantelhardtKRHB01}.  More importantly,
when the record of the whole community is shuffled but preserving the
inter-event distribution, the asymptotic scaling is $F\sim (\Delta
t)^{1/2}$. That is, in contrast to the result for individual activity,
when the shuffle the signal of the whole community, we obtain the
uncorrelated exponent: $H\simeq 0.5$ (dashed lines in
Fig.~\ref{fig:pokrecall3600seadfa}).  The fact that the correlations
vanish ($H=0.9$ $\rightarrow$ $H=0.5$) when the data is shuffled
indicates that the long-term correlations found in the activity of the
community as a whole are not due to Levy correlations. Instead,
correlations in the whole community are ``true correlations''
appearing as a manifestation of collective behavior of the scale of
the entire community.

Another surprise appears when we calculate the distribution of
inter-event times for the whole community. Here we define inter-event
the time between the sending of two consecutive messages of any member
in the community.  This contrasts with the same study done at the
single user level (Fig.~\ref{fig:oc2iet}) when inter-event is defined
as the time between two events of the same user.  In a sense,
$P({\rm d}t)$ for the entire community captures the collective
behavior emerging from the entire community as the information travels
through the whole network.

%Finally, we analyze the distribution of inter-event times. 
%Again, of the entire community, i.e. the intervals between any 
%two subsequent messages sent in the community 
%(we do not perform any deseasoning).
In Fig.~\ref{fig:pokietalldathst} the resulting probability density is
displayed.  We find a plateau up to $50$~seconds followed by a
power-law decay according to Eq.~(\ref{eq:pdtdelta}) with $\mu\approx
2.25$.
%This exponent implies $H_\mu\approx 0.88$, which is close to what we
%find in Fig.~\ref{fig:pokrecall3600seadfa}.  This agreement indicates
%Levy correlations.
Thus, the distribution of inter-event activity of the community as a
whole is also a Levy type like the single user activity, albeit with a
larger exponent.  Such a larger exponent reflects the fact that
$P({\rm d}t)$ is narrower for the community than for the individuals,
as expected.

When we convert the exponent $\mu\approx 2.25$ to the $H_\mu$ through the
Levy distribution model, Eq.~(\ref{eq:hdelta}), we find $H_\mu\approx 0.88$.
Thus surprisingly, Eq.~(\ref{eq:hdelta}) may also explain the
persistence as in the individual activity.  However, the main evidence
of Fig.~\ref{fig:pokrecall3600seadfa}, that is, the fact that the
correlations vanish when we shuffle the data, probe that, even if
Eq.~(\ref{eq:hdelta}) provides a good estimation of $H$, the long-term
correlations are due to 'real' correlations and are not an artifact of
the integration of a Levy type activity with DFA.

The long-term correlations found in the behavior of the entire
community is more understandable than in the activity of single
members, since the activity of the community is based on the
communication patterns of the messages and information flowing through
the whole system. The existence of $H\approx 0.9$ at the whole level and the
indications that the correlations are real ones is an interesting
instance of the emergence of critical behavior in the collective
dynamics of the system as a whole.

We conclude that while at the individual level we find Levy
correlations, the activity of the whole community comprises 'real'
correlations, which is due to the (possibly correlated) superposition
of the individuals activity into a collective self-organized
information flow in the system. Such a behavior is reminiscent of
critical systems in phase transitions.

\section*{Discussion}
\label{sec:conclusions}

We have studied the timing of communication in a social online
community and find long-term persistence in the activity of sending
messages at the single user level and the whole community
level. Furthermore, we have addressed the question of the origin of
these long-term correlations and whether these are Levy type or 'real'
correlations.  While in the case of Levy type correlations the
inter-event times need to be power-law distributed, 'real' long-term
correlations are independent of the distributions, since they are due
to interdependencies in the activities.  

Our work, then, still leaves unanswered the question of the cause of
the long-term persistence in the communication patterns at the whole
level. One possibility is that the temporal correlations are related
to correlations in the network structure
\cite{KentsisA2006,RybskiRK2010}.  The persistence could also be due
to social effects, i.e. the dynamics in the social network
\cite{StehleBB2010} induces persistent fluctuations, such as cascades.
An example could be that a group of friends tries to make an
appointment and therefore sends many subsequent messages in a
relatively short time \cite{PallaBV2007}.  After agreeing, the
communication activity among the group drops.  The activity patterns
of individuals could be understood as a superposition of many such
cascades.  On the other hand, it could be purely due to a state of
mind \cite{AllegriniMBFGGWP2009}, solipsistic, emerging from moods.
More research is needed to thoroughly understand the interesting properties
of human activity and its motives.

In conclusion, we have determined $3$ exponents to characterize
communication activity: (i) $H$, the fluctuation exponent of the
original data, (ii) $H_{\rm shuf}$, the fluctuation exponent when the
data is shuffled preserving the inter-event times, (iii) $H_\mu$, the
fluctuation exponent which is expected from power-law distributed
inter-event times.  We find that $H\approx H_{\rm shuf}\approx
H_\mu\approx 0.9$ which supports the hypothesis of Levy correlations
in the single user activity, while we find $H\approx 0.9 \ne H_{\rm
  shuf}\approx 0.5$ for the collective behavior of the whole community
revealing non-trivial long-term correlations and self-organization at
the level of the whole system.

We should mention a third scenario which we leave for future work.  It
is possible that the correlations comprise more complex features.  It
has been shown that nonlinear correlations in multifractal data sets
lead to power-law distributed inter-event times (of peaks over
threshold) \cite{BogachevEB2007}.  In fact, the authors of
\cite{BogachevEB2007} find in their Fig.~1(c) a similar dependence of
$\mu$ on the total number of events as we do for $H_\mu$ in our
Fig.~\ref{fig:dfaaveallbinfitiet}.  Additional analysis is needed to
fully characterize the multifractal properties
\cite{KantelhardtZKHBS02,KantelhardtRZBKLHB03,KantelhardtKBRBBH2005}
of communication activity via e-mails or messages in online
communities.

\vspace{1cm}

%\bibliographystyle{unsrt}
%\bibliography{rybski.bib}% Produces the bibliography via BibTeX.

\begin{thebibliography}{10}

\bibitem{BrockmannHG2006}
D.~Brockmann, L.~Hufnagel, and T.~Geisel.
\newblock The scaling laws of human travel.
\newblock {\em Nature}, 439(7075):462--465, 2006.

\bibitem{GonzalezHB2008}
M.~C. Gonzalez, C.~A. Hidalgo, and A.-L. Barab{\'a}si.
\newblock Understanding individual human mobility patterns.
\newblock {\em Nature}, 453(7196):779--782, 2008.

\bibitem{BarabasiAL2005}
A.-L. Barab{\'a}si.
\newblock The origin of bursts and heavy tails in human dynamics.
\newblock {\em Nature}, 435(7039):207--211, 2005.

\bibitem{MalmgrenSMA2008}
R.~D. Malmgren, D.~B. Stouffer, A.~E. Motter, and L.~A.~N. Amaral.
\newblock A poissonian explanation for heavy tails in e-mail communication.
\newblock {\em Proc. Nat. Acad. Sci. U.S.A.}, 105(47):18153--18158, 2008.

\bibitem{RybskiBHLM2009}
D.~Rybski, S.~V. Buldyrev, S.~Havlin, F.~Liljeros, and H.~A. Makse.
\newblock Scaling laws of human interaction activity.
\newblock {\em Proc. Nat. Acad. Sci. U.S.A.}, 106(31):12640--12645, 2009.

\bibitem{KantelhardtJW2010}
J.~W. Kantelhardt.
\newblock {\em Encyclopedia of Complexity and System Science}, chapter entry
  00620: Fractal and Multifractal Time Series.
\newblock Springer, 2009.

\bibitem{makse-pre}
H.~A. Makse, S.~Havlin, M.~Schwartz, and H.~E. Stanley.
\newblock Method for generating long-range correlations for large systems.
\newblock {\em Phys. Rev. E}, 53:5445--5449, 1996.

\bibitem{PengBGHSSS1992}
C.-K. Peng, S.~V. Buldyrev, A.~L. Goldberger, S.~Havlin, F.~Sciortino,
  M.~Simons, and H.~E. Stanley.
\newblock Long-range correlations in nucleotide sequences.
\newblock {\em Nature}, 356(6365):168--170, 1992.

\bibitem{PengMHHSG1993}
C.-K. Peng, J.~Mietus, J.~M. Hausdorff, S.~Havlin, H.~E. Stanley, and A.~L.
  Goldberger.
\newblock Long-range anticorrelations and non-gaussian behavior of the
  heartbeat.
\newblock {\em Phys. Rev. Lett.}, 70(9):1343--1346, 1993.

\bibitem{KoscielnyBundeBHRGS98}
E.~Koscielny-Bunde, A.~Bunde, S.~Havlin, H.~E. Roman, Y.~Goldreich, and H.-J.
  Schellnhuber.
\newblock Indication of a universal persistence law governing atmospheric
  variability.
\newblock {\em Phys. Rev. Lett.}, 81(3):729--732, 1998.

\bibitem{TadakiKNNSSY2006}
S.~Tadaki, M.~Kikuchi, A.~Nakayama, K.~Nishinari, A.~Shibata, Y.~Sugiyama, and
  S.~Yukawa.
\newblock Power-law fluctuation in expressway traffic flow: Detrended
  fluctuation analysis.
\newblock {\em J. Phys. Soc. Jpn.}, 75(3):034002, 2006.

\bibitem{XiaoYanZHM2007}
Z.~Xiao-Yan, L.~Zong-Hua, and T.~Ming.
\newblock Detrended fluctuation analysis of traffic data.
\newblock {\em Chin. Phys. Lett.}, 24(7):2142--2145, 2007.

\bibitem{KaempfTKM2011}
M.~K\"ampf, S.~Tismer, J.~W. Kantelhardt, and L.~Muchnik.
\newblock Burst event and return interval statistics in wikipedia access and
  edit data.
\newblock {\em submitted}, 2011.

\bibitem{LelandTWW1994}
W.~E. Leland, M.~S. Taqqu, W.~Willinger, and D.~V. Wilson.
\newblock On the self-similar nature of ethernet traffic (extended version).
\newblock {\em IEEE/ACM Trans. Networking}, 2(1):1--15, 1994.

\bibitem{LiuGCMPS1999}
Y.~Liu, P.~Gopikrishnan, P.~Cizeau, M.~Meyer, C.-K. Peng, and H.~E. Stanley.
\newblock Statistical properties of the volatility of price fluctuations.
\newblock {\em Phys. Rev. E}, 60(2):1390--1400, 1999.

\bibitem{MantegnaS1999}
R.~N. Mantegna and H.~E. Stanley.
\newblock {\em An Introduction to Econophysics: Correlations and Complexity in
  Finance}.
\newblock Cambridge University Press, Cambridge, 1999.

\bibitem{LuxA2002}
F.~Lux and M.~Ausloos.
\newblock {\em The Science of Disasters}, chapter 13. Market Fluctuations I:
  Scaling, Multiscaling, and Their Possible Origins, pages 373--409.
\newblock Springer-Verlag, Berlin, 2002.

\bibitem{SchenkelZZ1993}
A.~Schenkel, J.~Zhang, and Y.-C. Zhang.
\newblock Long range correlations in human writings.
\newblock {\em Fractals}, 1(1):47--57, 1993.

\bibitem{KosmidisKA2006}
K.~Kosmidis, A.~Kalampokis, and P.~Argyrakis.
\newblock Language time series analysis.
\newblock {\em Physica A}, 370(2):808--816, 2006.

\bibitem{IvanovBAHFBSG1999}
P.~Ch. Ivanov, A.~Bunde, L.~A.~N. Amaral, S.~Havlin, J.~Fritsch-Yelle, R.~M.
  Baevsky, H.~E. Stanley, and A.~L. Goldberger.
\newblock Sleep-wake differences in scaling behavior of the human heartbeat:
  Analysis of terrestrial and long-term space flight data.
\newblock {\em EPL}, 48(5):594--600, 1999.

\bibitem{BundeHKPPV00}
A.~Bunde, S.~Havlin, J.~W. Kantelhardt, T.~Penzel, J.-H. Peter, and K.~Voigt.
\newblock Correlated and uncorrelated regions in heart-rate fluctuations during
  sleep.
\newblock {\em Phys. Rev. Lett.}, 85(17):3736--3739, 2000.

\bibitem{LinkenkaerHansenNPI2001}
K.~Linkenkaer-Hansen, V.~V. Nikouline, J.~M. Palva, and R.~J. Ilmoniemi.
\newblock Long-range temporal correlations and scaling behavior in human brain
  oscillations.
\newblock {\em J. Neurosci.}, 21(4):1370--1377, 2001.

\bibitem{AllegriniMBFGGWP2009}
P.~Allegrini, D.~Menicucci, R.~Bedini, L.~Fronzoni, A.~Gemignani, P.~Grigolini,
  B.~J. West, and P.~Paradisi.
\newblock Spontaneous brain activity as a source of ideal $ 1/f $ noise.
\newblock {\em Phys. Rev. E}, 80(6):061914, 2009.

\bibitem{brain}
L.~K. Gallos, H.~A. Makse, and M.~Sigman.
\newblock A small-world of weak ties provides optimal global integration of
  self-similar modules in functional brain networks.
\newblock {\em Proc. Nat. Acad. Sci. USA}, 109:2825--2830, 2012.

\bibitem{IvanovHHSS2007}
P.~Ch. Ivanov, K.~Hu, M.~F. Hilton, S.~A. Shea, and H.~E. Stanley.
\newblock Endogenous circadian rhythm in human motor activity uncoupled from
  circadian influences on cardiac dynamics.
\newblock {\em Proc. Nat. Acad. Sci. U.S.A.}, 104(52):20702--20707, 2007.

\bibitem{makse-city}
H.~A. Makse, S.~Havlin, and H.~E. Stanley.
\newblock Modelling urban growth patterns.
\newblock {\em Nature}, 377:608--612, 1995.

\bibitem{makse-perco}
H.~A. Makse, J.~S. Andrade, M.~Batty, S.~Havlin, and H.~E. Stanley.
\newblock Modeling urban growth patterns with correlated percolation.
\newblock {\em Phys. Rev. E}, 58:7054--7062, 1998.

\bibitem{rozenfeld-pnas}
H.~D. Rozenfeld, D.~Rybski, J.~S. Andrade~Jr., M.~Batty, H.~E. Stanley, and
  H.~A. Makse.
\newblock Laws of population growth.
\newblock {\em Proc. Nat. Acad. Sci. USA}, 105:18702--18707, 2008.

\bibitem{rozenfeld-aer}
H.~D. Rozenfeld, D.~Rybski, and H.~A. Gabaix, X.~Makse.
\newblock The area and population of cities: New insights from a different
  perspective on cities.
\newblock {\em American Economic Review}, 101:560--580, 2011.

\bibitem{galvao}
G.~Galvao, J.~G.~V. Miranda, R.~F.~S. Andrade, J.~S. Andrade~Jr., L.~K. Gallos,
  and H.~A. Makse.
\newblock Modularity map of the network of humn cell differentiation.
\newblock {\em Proc. Nat. Acad. Sci. USA}, 107:5750--5755, 2010.

\bibitem{obesity}
L.~K. Gallos, P.~Barttfeld, S.~Havlin, M.~Sigman, and H.~A. Makse.
\newblock Collective behavior in the spatial spreading of obesity.
\newblock {\em Sci. Rep.}, 2012.

\bibitem{ShlesingerFK1987}
M.~F. Shlesinger, B.~J. West, and J.~Klafter.
\newblock L\'evy dynamics of enhanced diffusion: Application to turbulence.
\newblock {\em Phys. Rev. Lett.}, 58(11):1100--1103, 1987.

\bibitem{HolmeP2003}
P.~Holme.
\newblock Network dynamics of ongoing social relationships.
\newblock {\em EPL}, 64(3):427--433, 2003.

\bibitem{HolmeLEK2003}
P.~Holme, F.~Liljeros, C.~R. Edling, and B.~J. Kim.
\newblock Network bipartivity.
\newblock {\em Phys. Rev. E}, 68(5):056107, 2003.

\bibitem{HolmeEL2004}
P.~Holme, C.~R. Edling, and F.~Liljeros.
\newblock Structure and time evolution of an internet dating community.
\newblock {\em Soc. Networks}, 26(2):155--174, 2004.

\bibitem{RybskiBHLM5b}
D.~Rybski, S.~V. Buldyrev, S.~Havlin, F.~Liljeros, and H.~A. Makse.
\newblock Communication activity in social networks: growth and correlations.
\newblock {\em Eur. Phys. J. B}, 84(1):147--159, 2011.

\bibitem{GallosRLHM2011}
L.~K. Gallos, D.~Rybski, F.~Liljeros, S.~Havlin, and H.~A. Makse.
\newblock How people interact in evolving online affiliation networks.
\newblock {\em submitted}, 2011.

\bibitem{PengBHSSG94}
C.-K. Peng, S.~V. Buldyrev, S.~Havlin, M.~Simons, H.~E. Stanley, and A.~L.
  Goldberger.
\newblock Mosaic organization of dna nucleotides.
\newblock {\em Phys. Rev. E}, 49(2):1685--1689, 1994.

\bibitem{KantelhardtKRHB01}
J.~W. Kantelhardt, E.~Koscielny-Bunde, H.~H.~A. Rego, S.~Havlin, and A.~Bunde.
\newblock Detecting long-range correlations with detrended fluctuation
  analysis.
\newblock {\em Physica A}, 295(3--4):441--454, 2001.

\bibitem{GolderWH2006}
S.~Golder, D.~M. Wilkinson, and B.~A. Huberman.
\newblock Rhythms of social interaction: messaging within a massive online
  network.
\newblock {\em online-arXiv}, arXiv:cs/0611137v1 [cs.CY], 2006.

\bibitem{LeskovecH2008}
J.~Leskovec and E.~Horvitz.
\newblock Planetary-scale views on an instant-messaging network.
\newblock {\em online-arXiv}, arXiv:0803.0939v1 [physics.soc-ph], 2008.

\bibitem{MalmgrenSCA2009}
R.~D. Malmgren, D.~B. Stouffer, A.~S. L.~O. Campanharo, and L.~A~.N Amaral.
\newblock On universality in human correspondence activity.
\newblock {\em Science}, 325(5948):1696--1700, 2009.

\bibitem{EislerK2006}
Z.~Eisler and J.~Kert\'esz.
\newblock Scaling theory of temporal correlations and size-dependent
  fluctuations in the traded value of stocks.
\newblock {\em Phys. Rev. E}, 73(4):046109, 2006.

\bibitem{EislerBK2008}
Z.~Eisler, I.~Bartos, and J.~Kert\'esz.
\newblock Fluctuation scaling in complex systems: Taylor's law and beyond.
\newblock {\em Adv. Phys.}, 57(1):89--142, 2008.

\bibitem{JohansenA2004}
A.~Johansen.
\newblock Probing human response times.
\newblock {\em Physica A}, 338(1--2):286--291, 2004.

\bibitem{JohansenA2006}
A.~Johansen.
\newblock Comment on {A}.-{L}. {B}arabasi, {N}ature {\bf 435} 207-211 (2005).
\newblock {\em online-arXiv}, arXiv:physics/0602029v1 [physics.soc-ph], 2006.

\bibitem{VazquezA2005}
A.~V\'azquez.
\newblock Exact results for the barabasi model of human dynamics.
\newblock {\em Phys. Rev. Lett.}, 95(24):248701, 2005.

\bibitem{VazquezODGKB2006}
A.~V\'azquez, J.~G. Oliveira, Z.~Dezs\"o, K.~I. Goh, I.~Kondor, and A.-L.
  Barab\'asi.
\newblock Modeling bursts and heavy tails in human dynamics.
\newblock {\em Phys. Rev. E}, 73(3):036127, 2006.

\bibitem{StoufferMA2005}
D.~B. Stouffer, R.~D. Malmgren, and L.~A.~N. Amaral.
\newblock Comment on "{T}he origin of bursts and heavy tails in human dynamics"
  by {B}arabasi, {N}ature {\bf 435}, 207 (2005).
\newblock {\em online-arXiv}, arXiv:physics/0510216v1 [physics.data-an], 2005.

\bibitem{BarabasiGV2005}
A.-L. Barab\'asi, K.-I. Goh, and A.~Vazquez.
\newblock Reply to comment on "the origin of bursts and heavy tails in human
  dynamics".
\newblock {\em online-arXiv}, arXiv:physics/0511186v1 [physics.data-an], 2005.

\bibitem{NewmanM2005}
M.~E.~J. Newman.
\newblock Power laws, {P}areto distributions and {Z}ipf's law.
\newblock {\em Contemp. Phys.}, 46(5):323--351, 2005.

\bibitem{GabaixI2006}
X.~Gabaix and R.~Ibragimov.
\newblock Log(rank-1/2): a simple way to improve the ols estimation of tail
  exponents.
\newblock Discussion Paper 2106 (26), Harvard Institute of Economic Research,
  Cambridge, Massachusetts, February 2006.

\bibitem{ClausetRSN2007}
A.~Clauset, C.~R. Shalizi, and M.~E.~J. Newman.
\newblock Power-law distributions in empirical data.
\newblock {\em SIAM Rev.}, 51(4):661--703, 2009.

\bibitem{MalevergnePS2009}
Y.~Malevergne, V.~Pisarenko, and D.~Sornette.
\newblock Gibrat's law for cities: uniformly most powerful unbiased test of the
  {P}areto against the lognormal.
\newblock {\em online-arXiv}, arXiv:0909.1281v1, 2009.

\bibitem{GohB2008}
K.-I. Goh and A.-L. Barab{\'a}si.
\newblock Burstiness and memory in complex systems.
\newblock {\em EPL}, 81(4):48002, 2008.

\bibitem{EichnerKBH2007}
J.~F. Eichner, J.~W. Kantelhardt, A.~Bunde, and S.~Havlin.
\newblock Statistics of return intervals in long-term correlated records.
\newblock {\em Phys. Rev. E}, 75(1):011128, 2007.

\bibitem{LennartzB2009}
S.~Lennartz and A.~Bunde.
\newblock Eliminating finite-size effects and detecting the amount of white
  noise in short records with long-term memory.
\newblock {\em Phys. Rev. E}, 79(6):066101, 2009.

\bibitem{GersteinM1964}
G.~L. Gerstein and B.~Mandelbrot.
\newblock Random walk models for spike activity of single neuron.
\newblock {\em Biophys. J.}, 4(1P1):41--68, 1964.

\bibitem{BuldyrevGHPSS1993}
S.~V. Buldyrev, A.~L. Goldberger, S.~Havlin, C.-K. Peng, M.~Simons, and H.~E.
  Stanley.
\newblock Generalized {L}\'evy-walk model for {DNA} nucleotide sequences.
\newblock {\em Phys. Rev. E}, 47(6):4514--4523, 1993.

\bibitem{BuldeyrevSV2010}
S.~V. Buldyrev.
\newblock {\em Encyclopedia of Complexity and System Science}, volume Fractals
  and multifractals, chapter Fractals in Biology.
\newblock Springer, 2010.

\bibitem{BundeEKH2005}
A.~Bunde, J.~F. Eichner, J.~W. Kantelhardt, and S.~Havlin.
\newblock Long-term memory: A natural mechanism for the clustering of extreme
  events and anomalous residual times in climate records.
\newblock {\em Phys. Rev. Lett.}, 94(4):048701, 2005.

\bibitem{OliveiraB2005}
J.~G. Oliveira and A.-L. Barab{\'a}si.
\newblock {D}arwin and {E}instein correspondence patterns.
\newblock {\em Nature}, 437(7063):1251--1251, 2005.

\bibitem{ThurnerLFHFT1997}
S.~Thurner, S.~B. Lowen, M.~C. Feurstein, C.~Heneghan, H.~G. Feichtinger, and
  M.~C. Teich.
\newblock Analysis, synthesis, and estimation of fractal-rate stochastic point
  processes.
\newblock {\em Fractals}, 5(4):565--595, 1997.

\bibitem{KentsisA2006}
A.~Kentsis.
\newblock Mechanisms and models of human dynamics.
\newblock {\em Nature}, 441(7092):E5--E5, 2006.

\bibitem{RybskiRK2010}
D.~Rybski, H.~D. Rozenfeld, and J.~P. Kropp.
\newblock Quantifying long-range correlations in complex networks beyond
  nearest neighbors.
\newblock {\em EPL}, 90(2):28002, 2010.

\bibitem{StehleBB2010}
J.~Stehle, A.~Barrat, and G.~Bianconi.
\newblock Dynamical and bursty interactions in social networks.
\newblock {\em Phys. Rev. E}, 81(3):035101, 2010.

\bibitem{PallaBV2007}
G.~Palla, A.-L. Barab{\'a}si, and T.~Vicsek.
\newblock Quantifying social group evolution.
\newblock {\em Nature}, 446(7136):664--667, 2007.

\bibitem{BogachevEB2007}
M.~I. Bogachev, J.~F. Eichner, and A.~Bunde.
\newblock Effect of nonlinear correlations on the statistics of return
  intervals in multifractal data sets.
\newblock {\em Phys. Rev. Lett.}, 99(24):240601, 2007.

\bibitem{KantelhardtZKHBS02}
J.~W. Kantelhardt, S.~A. Zschiegner, E.~Koscielny-Bunde, S.~Havlin, A.~Bunde,
  and H.~E. Stanley.
\newblock Multifractal detrended fluctuation analysis of nonstationary time
  series.
\newblock {\em Physica A}, 316(1--4):87--114, 2002.

\bibitem{KantelhardtRZBKLHB03}
J.~W. Kantelhardt, D.~Rybski, S.~A. Zschiegner, P.~Braun, E.~Koscielny-Bunde,
  V.~Livina, S.~Havlin, and A.~Bunde.
\newblock Multifractality of river runoff and precipitation: comparison of
  fluctuation analysis and wavelet methods.
\newblock {\em Physica A}, 330(1--2):240--245, 2003.

\bibitem{KantelhardtKBRBBH2005}
J.~W. Kantelhardt, E.~Koscielny-Bunde, D.~Rybski, P.~Braun, A.~Bunde, and
  S.~Havlin.
\newblock Long-term persistence and multifractality of precipitation and river
  runoff records.
\newblock {\em J. Geophys. Res.-Atmos.}, 111(D1):D01106, 2006.

\end{thebibliography}

\section*{Acknowledgments}
We thank C. Briscoe, J.F. Eichner, L.K. Gallos, and H.D. Rozenfeld for
useful discussions.  This work was supported by National Science
Foundation Grants NSF-SES-0624116 and NSF-EF-0827508 and ARL.
F.L. acknowledges financial support from The Swedish Bank Tercentenary
Foundation.  S.H. thanks the European EPIWORK project, the Israel
Science Foundation, ONR and DTRA for financial support.

\section*{Author contributions}
All authors contributed equally to the work presented in this paper.

\section*{Additional information}
The authors declare no competing financial interests.

\newpage

\begin{figure}
\centerline{\resizebox{15.0cm}{!} { \includegraphics{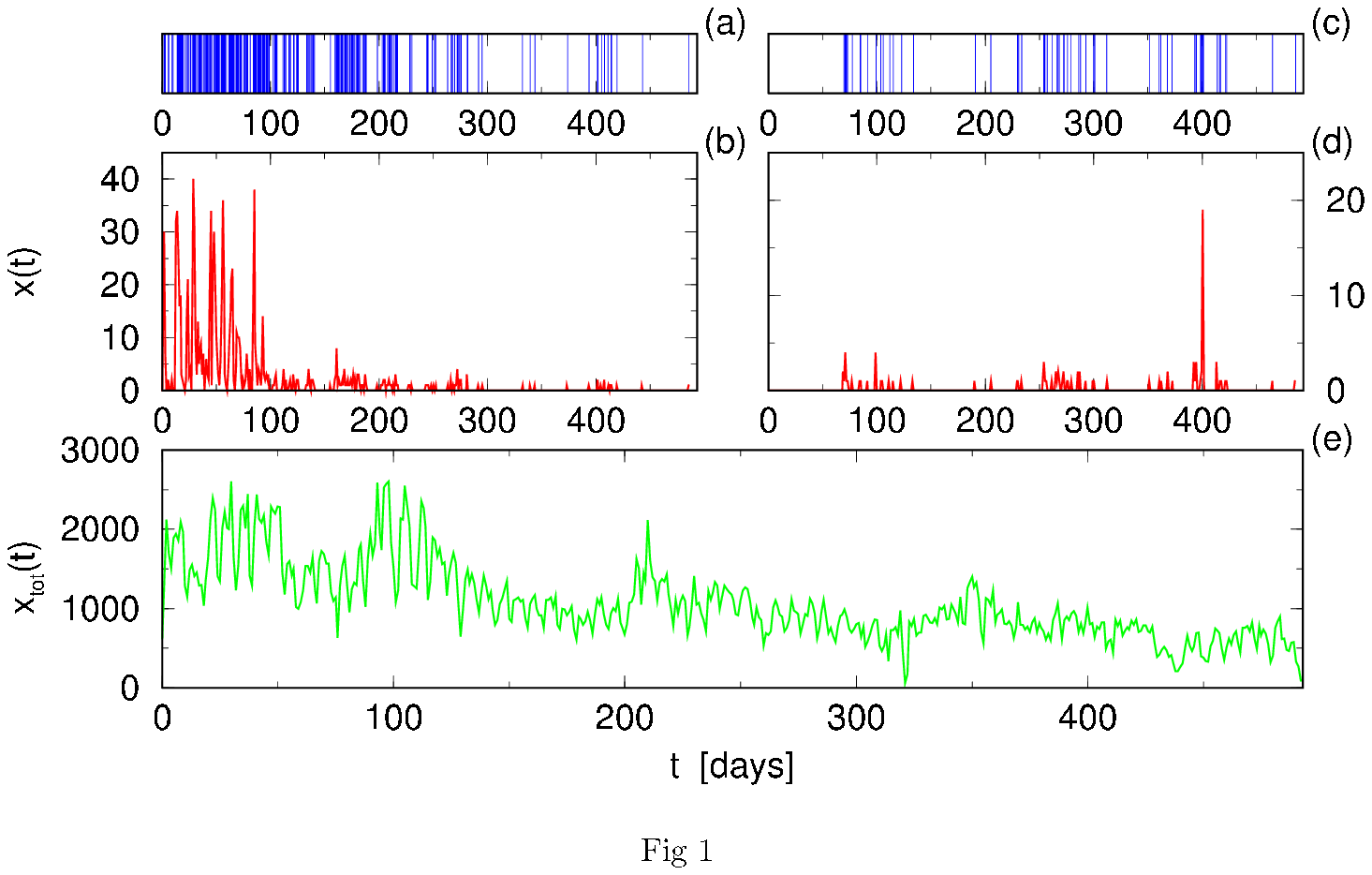}}}
\caption{}
\label{fig:pokillu}
\end{figure}

FIG. \ref{fig:pokillu}. Examples of activity of sending messages and
overall activity in POK.  The vertical lines in (a) and (c) represent
the instants when the messages have been sent by two arbitrary
members.  The panels (b) and (d) show the records of number of
messages per day, $x(t)$, of the same two members.  The record of the
total number of messages sent by all members per day within POK is
depicted in (e).  (a) and (b): member~326 ($M=1023$); (c) and (d):
member~9414 ($M=100$).

\begin{figure}
\centerline{\resizebox{15.0cm}{!} {
    \includegraphics{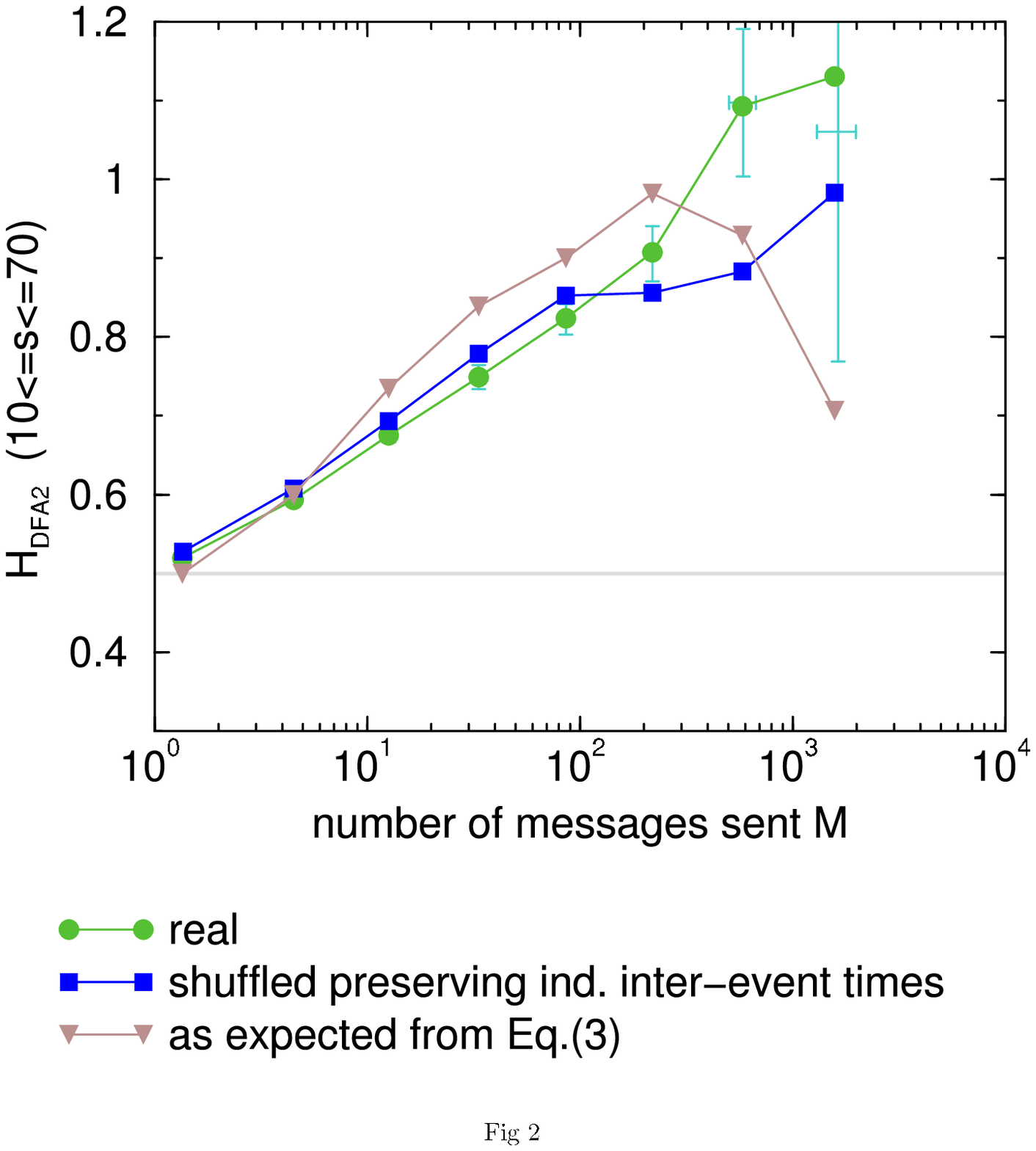} }}\caption{}
\label{fig:dfaaveallbinfitiet}
\end{figure}

FIG. \ref{fig:dfaaveallbinfitiet}.  Fluctuation exponents of the
communication activity sending messages.  The exponents are plotted as
a function of the activity level~$M$ (final number of messages) for
the original data (green circles), shuffled data preserving the
individual inter-event times (blue squares), as well as the exponents
expected from Eq.~(\ref{eq:hdelta}) (brown triangles down) and from
the distribution of inter-event times as characterized by power-law
fits to the curves of Fig.~\ref{fig:oc2iet} providing the
exponent~$\mu$.

\begin{figure}
\centerline{\resizebox{15.0cm}{!} { \includegraphics{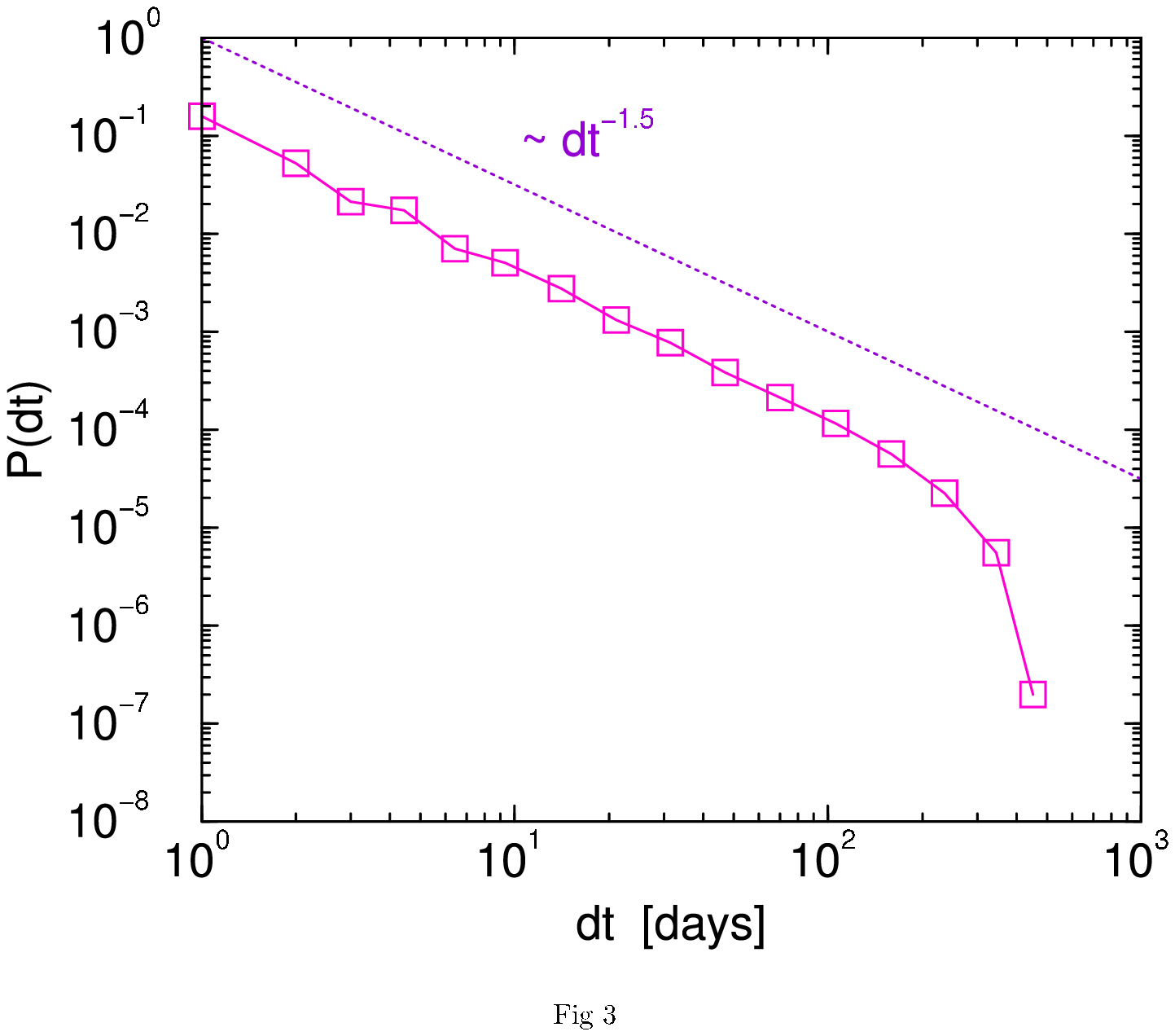} }}
\caption{}
\label{fig:ietall}
\end{figure}

\begin{figure}
\centerline{\resizebox{15.0cm}{!} { \includegraphics{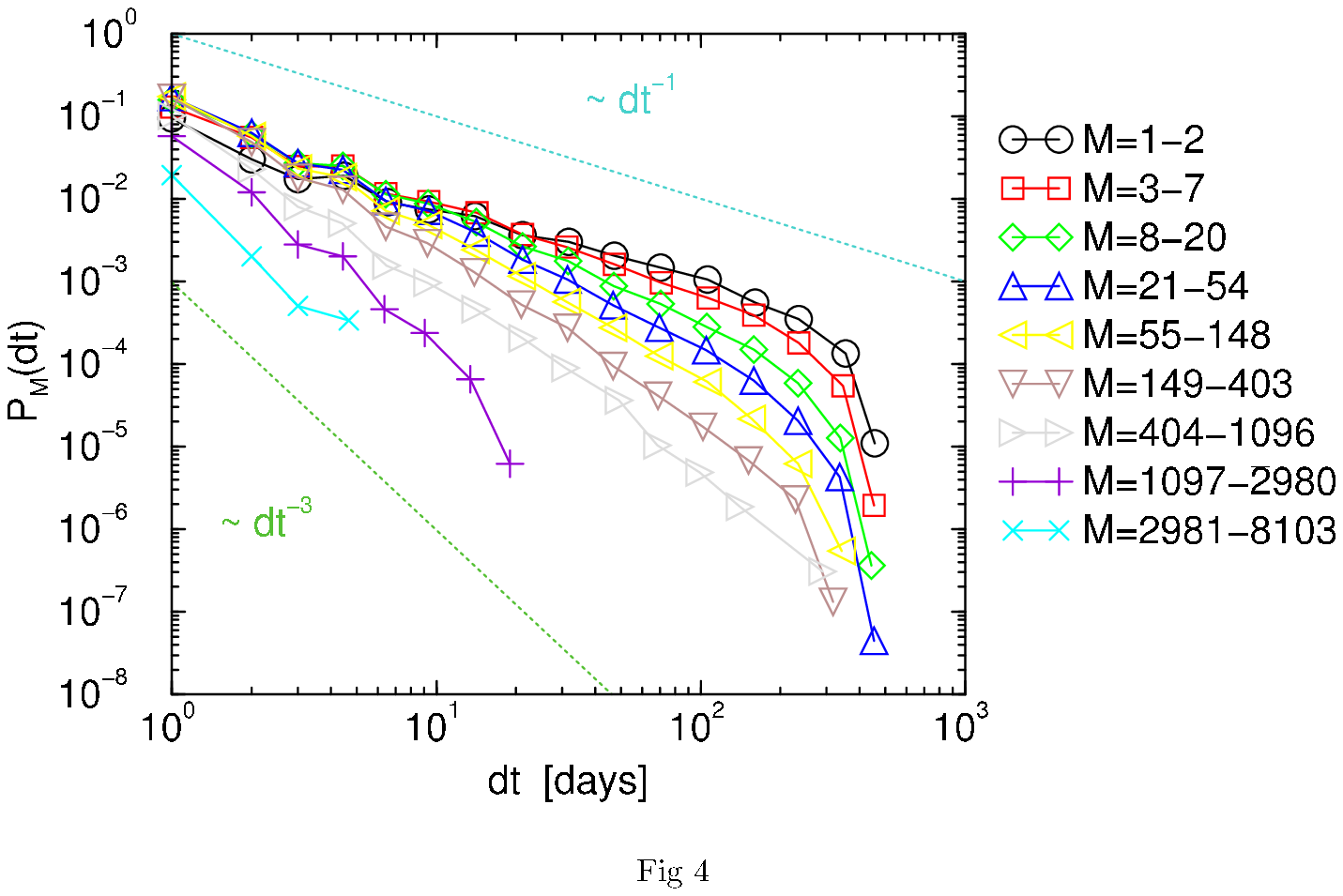} }}
\caption{}
\label{fig:oc2iet}
\end{figure}

FIG. \ref{fig:ietall}.  Probability density of inter-event times
${\rm d}t$ between successive messages sent by a single member of POK,
in daily resolution.  The values are extracted by considering every
single individual sending messages in the period of data acquisition
and then joined from all members.  The dotted line in the top
corresponds to the exponent~$\mu=1.5$.

FIG. \ref{fig:oc2iet}.  Probability density of inter-event times
${\rm d}t$ between successive messages sent by all individual members
of POK in daily resolution.  The values are extracted for the
individuals and unified among members according to their activity
level~$M$.  The curve for the most active members is in the bottom,
while the one for the least active is in the top.  The dotted lines
correspond to the exponents $-1$ (top) and $-3$ (bottom).

\begin{figure}
\centerline{\resizebox{15.0cm}{!} { \includegraphics{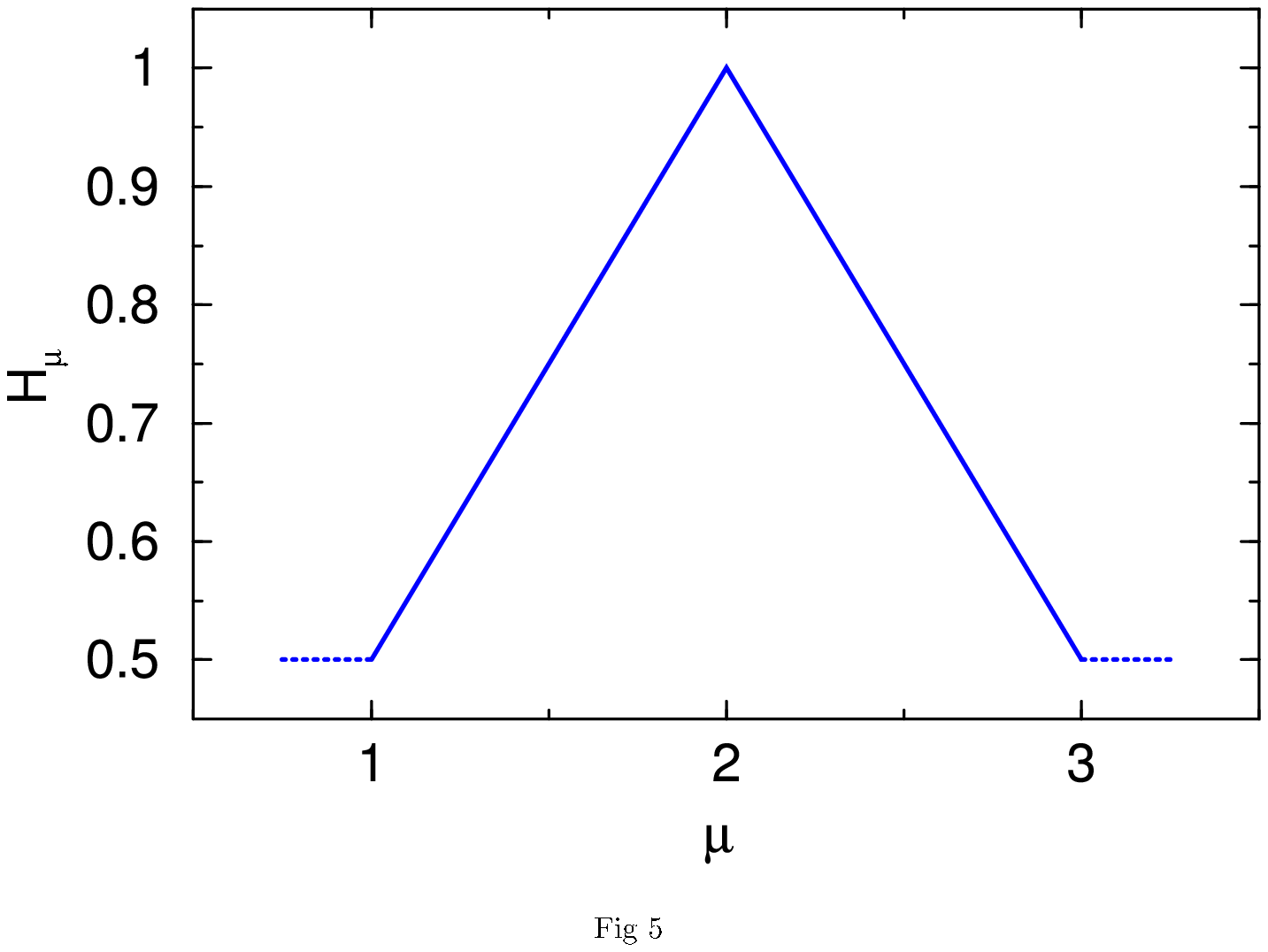}}}
\caption{}
\label{fig:prediction}
\end{figure}

FIG. \ref{fig:prediction}.  Levy correlations and persistence.
Theoretical relation between the inter-event time distribution
exponent~$\mu$ and the fluctuation exponent~$H_\mu$ according to
Eq.~(\ref{eq:hdelta}).

\begin{figure}
\centerline{\resizebox{15.0cm}{!} {
    \includegraphics{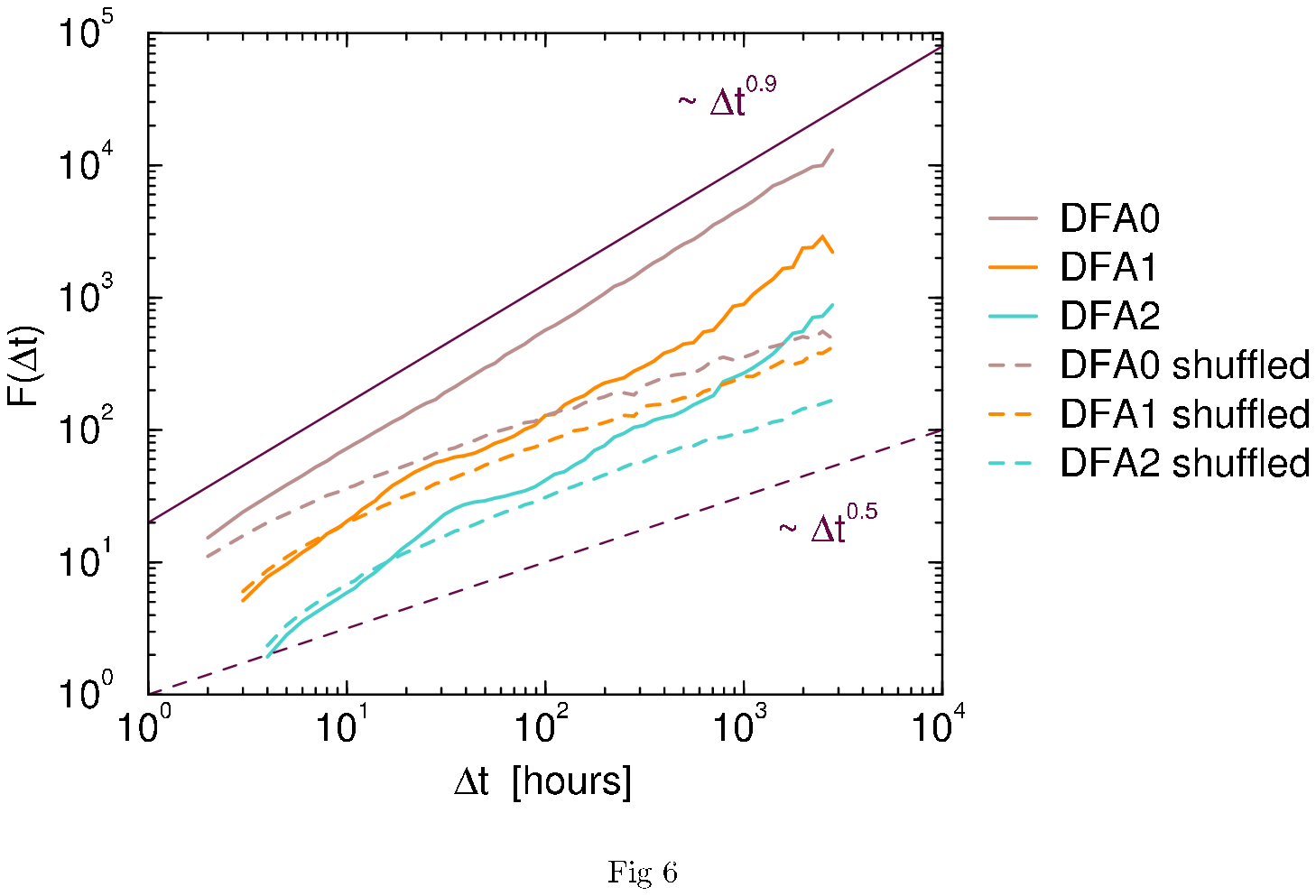}}}
\caption{}
\label{fig:pokrecall3600seadfa}
\end{figure}

FIG. \ref{fig:pokrecall3600seadfa}. Fluctuation function of the
record of messages sent by any member of POK.  The record is the same
as in Fig.~\ref{fig:pokillu}(e) but in hourly resolution.  Prior to
applying DFA, the record has been deseasoned according to $x_{\rm
  tot}(t) \rightarrow x_{\rm tot}(t)-\langle x_{\rm
  tot}\rangle_{t\%24}$.  The different curves differ in the DFA-order
(DFA0-DFA2, from top to bottom), which determines the capability of
detrending.  DFA2 eliminates linear trends in $x_{\rm tot}(t)$
\cite{KantelhardtKRHB01}.  The dotted line in the bottom corresponds
to a power-law with exponent $H=0.5$ and serves as guide to the eye
while the continuous line at the top represents $H=0.9$.

\begin{figure}
\centerline{\resizebox{15.0cm}{!} {
    \includegraphics{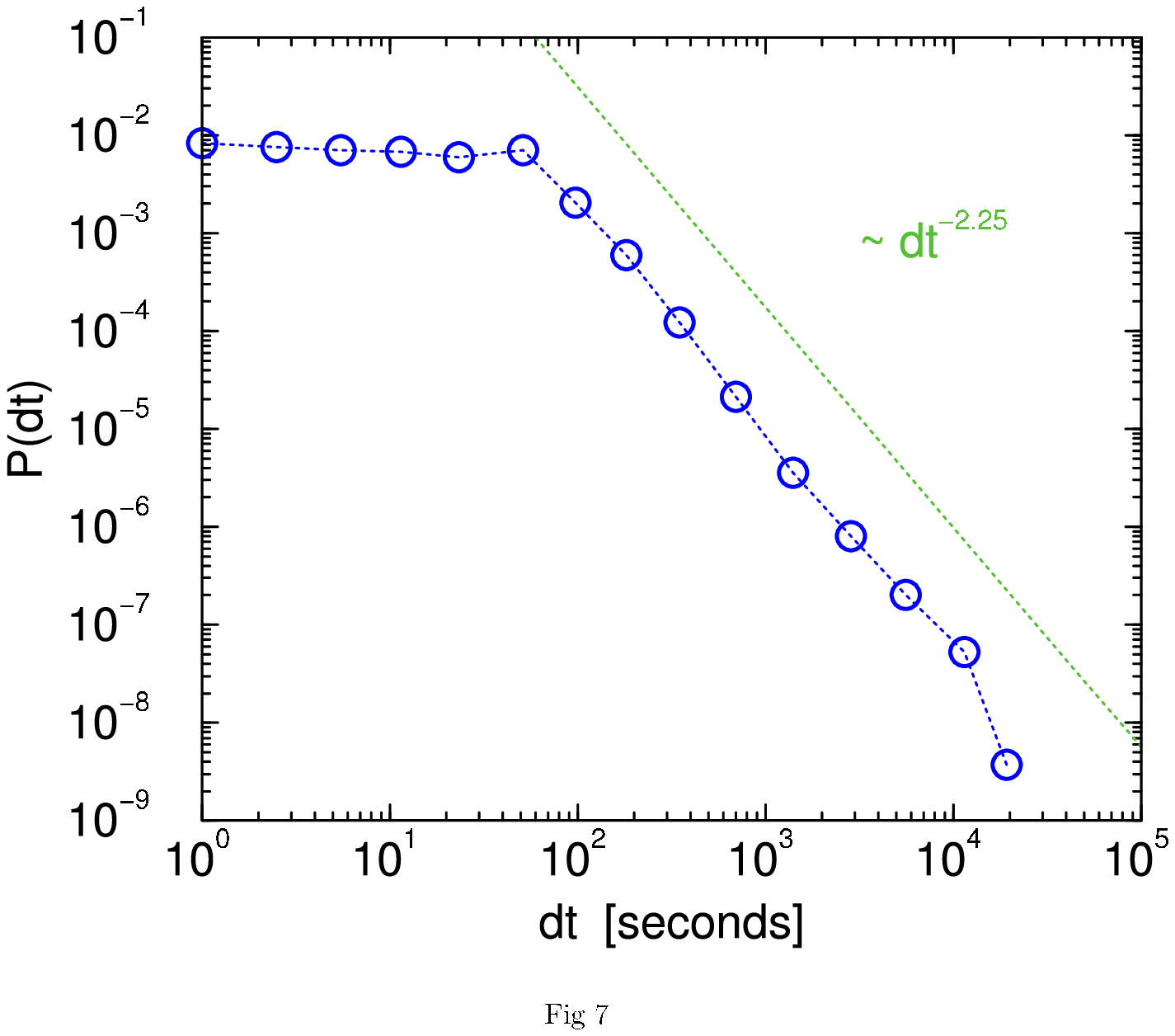}}}
\caption{}
\label{fig:pokietalldathst}
\end{figure}

FIG. \ref{fig:pokietalldathst}. Probability density of inter-event
times ${\rm d}t$ between successive messages sent by any member of POK
in seconds. The dotted straight line has corresponds to a power-law
with exponent $\mu=-2.25$ and serves as guide to the eye.

\newpage

\newpage

\newpage

\newpage
\newpage
\newpage
\newpage

\end{document}